\documentclass[12pt,a4paper]{article}
\usepackage{amsmath,amsthm}
\usepackage{amsmath,amssymb}
\usepackage{cases}

\numberwithin{equation}{section}

\newcommand {\Cg} {{\cal C}} 
\newcommand {\Eg} {{\cal E}} 

\newcommand {\Hg} {{\cal H}} 
\newcommand {\Ng} {{\cal N}} 

\newcommand {\R} {{\mathbb R}}

\newcommand {\C} {{\mathbb C}}

\newcommand {\Bb} {{\bf B}}

\newcommand {\p} {\psi}
\newcommand {\f} {\phi}
\newcommand {\ep} {\varepsilon}

\newcommand {\lam} {\lambda}
\newcommand {\m} {\mu}
\newcommand {\n} {\nu}
\newcommand {\s} {\sigma}

\newcommand {\w} {\omega}

\def\lbeq(#1){\label{eqn:#1}}
\def\refeq(#1){{\rm (\ref{eqn:#1})}}
\def\lbth(#1){\label{th:#1}}
\def\refth(#1){{\rm Theorem \ref{th:#1}}}
\def\refthb(#1){{\bf Theorem \ref{th:#1}}}
\def\refths(#1,#2){{\rm Theorems \ref{th:#1} and \ref{th:#2}}}
\def\lblm(#1){\label{lm:#1}}
\def\reflm(#1){{\rm Lemma \ref{lm:#1}}}
\def\reflmss(#1,#2,#3){{\rm Lemmas \ref{lm:#1}, \ref{lm:#2} and \ref{lm:#3}}}
\def\reflms(#1,#2){{\rm Lemmas \ref{lm:#1} and \ref{lm:#2}}}
\def\reflmb(#1){{\bf Lemma \ref{lm:#1}}}
\def\lbrm(#1){\label{rm:#1}}
\def\refrm(#1){{\rm Remark \ref{rm:#1}}}
\def\lbprp(#1){\label{prp:#1}}
\def\refprp(#1){{\rm Proposition \ref{prp:#1}}}
\def\lbass(#1){\label{ass:#1}}
\def\refass(#1){{\rm Assumption \ref{ass:#1}}}
\def\lbcor(#1){\label{cor:#1}}
\def\refcor(#1){{\rm Corollary \ref{cor:#1}}}

\newcommand {\bgth} {\begin{theorem}}
\newcommand {\edth} {\end{theorem}}
\newcommand {\bgprp} {\begin{proposition}}
\newcommand {\edprp} {\end{proposition}}
\newcommand {\bgdf} {\begin{definition}}
\newcommand {\eddf} {\end{definition}}
\newcommand {\bgass} {\begin{assumption}}
\newcommand {\edass} {\end{assumption}}
\newcommand {\bglm} {\begin{lemma}}
\newcommand {\edlm} {\end{lemma}}
\newcommand {\bgcor} {\begin{corollary}}
\newcommand {\edcor} {\end{corollary}}
\newcommand {\bgpf} {\begin{proof}}
\newcommand {\edpf} {\end{proof}}
\newcommand {\bgrm} {\begin{remark}}
\newcommand {\edrm} {\end{remark}}

\newcommand {\bqn} {\begin{equation}}
\newcommand {\eqn} {\end{equation}}

\newcommand {\ben} {\begin{enumerate}}
\newcommand {\een} {\end{enumerate}}
\newcommand {\ph} {\varphi}

\newcommand {\la} {\langle}
\newcommand {\ra} {\rangle}

\newcommand {\ax} {{\la x \ra}}
\newcommand {\ay} {{\la y \ra}}

\newcommand {\br} {\begin{array}}
\newcommand {\er} {\end{array}}
\newcommand {\lap} {\Delta}
\newcommand {\pa} {\partial}

\newtheorem{theorem}{Theorem}[section]
\newtheorem{lemma}[theorem]{Lemma}
\newtheorem{proposition}[theorem]{Proposition}
\newtheorem{definition}[theorem]{Definition}
\newtheorem{corollary}[theorem]{Corollary}

\newtheorem{assumption}[theorem]{Assumption}
\theoremstyle{definition}
\newtheorem{remark}[theorem]{Remark}

\newcommand {\absleq} {{\leq_{|\, \cdot\, |}\, }}

\title{On wave operators for Schr\"odinger operators 
with threshold singuralities in three dimensions}
\author{
K. Yajima\thanks{SISSA-International School for Advanced Studies, 
Via Bonomea, 34136 Trieste, Italy. On leave from 
Department of Mathematics, Gakushuin University, 
1-5-1 Mejiro, Toshima-ku, Tokyo 171-8588, Japan. 
Supported by JSPS research No.16K05242}}
\date{}
\begin{document}
\allowdisplaybreaks
\maketitle

\begin{abstract}
We show that wave operators for three dimensional 
Schr\"odinger operators $H=-\lap + V$ with threshold 
singularities are bounded in $L^1(\R^3)$ if and only 
if zero energy resonances are absent from $H$ and the 
existence of zero energy eigenfunctions does not destroy 
the $L^1$-boundedness of wave operators for $H$ 
with the regular threshold behavior. 
We also show in this case that they are bounded in 
$L^p(\R^3)$ for all $1\leq p \leq \infty$ if all zero 
energy eigenfunctions $\f(x)$ have vanishing first three 
moments: 
$\int_{\R^3} x^\alpha V(x)\f(x)dx=0$, $|\alpha|=0,1,2$.   
\end{abstract}

\section{Introduction} 

Let $H_0=-\lap$ be the free Schr\"odinger operator on the 
Hilbert space $\Hg=L^2(\R^m)$ with domain 
$D(H_0)=\{u \in \Hg \colon \pa^\alpha u \in \Hg, \ |\alpha|\leq 2\}$ 
and $H= -\lap + V$, $V$ being the multiplication 
with real measurable function $V(x)$ such that 
$|V(x)|\leq C\ax^{-\delta}$ for some $\delta>2$, 
$\ax=(1+|x|^2)^\frac12$. Then, it is well known that 
wave operators $W_\pm$ defined by the strong limits 
in $\Hg=L^2(\R^m)$: 
\begin{equation}\lbeq(defw)
W_\pm  = \lim_{t \to\pm\infty} e^{itH}e^{-itH_0}
\end{equation}
exist, they are unitary from  $\Hg$ to the absolutely 
continuous spectral subspace $\Hg_{ac}(H)$ of $\Hg$ for $H$ 
and enjoy the intertwining property: 
\bqn \lbeq(inter) 
f(H)P_{ac}(H)= W_{\pm} f(H_0) W_{\pm}^\ast 
\eqn 
for any Borel functions on $\R^{1}$, where 
$P_{ac}(H)$ is the orthogonal projection onto $\Hg_{ac}(H)$. 
The intertwining property reduces the mapping properties of 
$f(H)P_{ac}(H)$ to those of $f(H_0)$ provided that corresponding 
properties of $W_\pm$ are already established. 
For this reason the $L^p$-boundedness of $W_\pm$ 
has attracted various authors' interest and many 
important results are obtained (see \cite{Y-doc} and 
references therein). 

We recall here the results for three dimensions, 
restricting ourselves to the case $m=3$ in what follows. 
We omit $\R^3$ from $L^p(\R^3)$ and etc.  
We write $L^2_\s =L^2(\R^3, \ax^{2\s}dx)$ for $\s\in \R$ and 
\[
\la f, g\ra = \int_{\R^3} f(x) g(x) dx 
\]
whenever the right hand side makes sense. 
Define for $1/2<\s<\delta-1/2$ that 
\bqn \Ng=\{u \in L^2_{-\s} \colon u+ (-\lap)^{-1} Vu =0\} 
\lbeq(res) 
\eqn 
and 
\bqn 
\Eg=\{u \in L^2 \colon u+ (-\lap)^{-1} Vu =0\}. 
\lbeq(eigen)
\eqn 
The space $\Ng$ is independent of $\s$; all $u\in \Ng$ satisfy 
\[
-\lap u + V u=0; 
\]
$\Eg \subset \Ng$ is the zero energy eigenspace of $H$;  
$u \in \Ng$ belongs to $\Eg$ if and only if 
$\la V, u \ra =0$ and $\dim\Ng/\Eg \leq 1$.  
Functions $\f \in \Ng \setminus \Eg$ are called resonances. 
\ben 
\item[\rm (a)] If $\Ng=\{0\}$ then, $W_\pm$ are bounded 
for all $1\leq p \leq \infty$ (\cite{Y-d3}). 
\item[\rm (b)] If $\Ng\not=\{0\}$, then 
$W_\pm$ are in general bounded in $L^p$ for $1<p<3$. 
They are bounded in $L^p$ for all 
$1<p<\infty$ if and only if all $\f \in \Ng$ satisfy 
$\la V, x^\alpha \f\ra=0$ for $|\alpha|\leq 1$ (\cite{Y-doc}).   
\een

In this paper, being inspired by the approached employed by 
Goldberg and Green (\cite{GG}) for proving the $L^p$-boundedness 
of wave operators including $p=1$ for higher dimensional 
Schr\"odinger operators with threshold singularities, 
we prove the following theorem for the end point 
cases $p=1$ and $p=\infty$ for three dimensions which are missing 
from the results mentioned in (b). 
\bgth \lbth(1)
Suppose that $|V(x)|\leq C \ax^{-7-\ep}$. Then, 
$W_\pm$ are bounded in $L^1$ if and only if $\Ng=\Eg$. 
They are bounded also in $L^\infty$ if all $\f\in \Ng$ satisfy  
$\la V,  x^{\alpha}\f\ra$ for $|\alpha|\leq 2$.
\edth 
Incidentally, the method of the proof of the theorem may 
be used for the proof of the ``if'' part of results 
in (b) which is different from the one given in \cite{Y-doc}.  
We present it here, however, only for $1<p<3$ for avoiding too 
much repetition. 
We think that the assumption on $V$ of the theorem is unnecessarily 
too strong, however, we do not pursue better conditions here.

We refer readers more about the $L^p$ boundedness of wave 
operators to \cite{Y-doc} and the literature therein 
and jump into the proof of the theorem immediately. We recall 
that $\f \in \Ng$ satisfies 
\bqn \lbeq(fng)
\f(x) + \frac1{4\pi}\int_{\R^3}\frac{V(y)\f(y)}{|x-y|}dy = 0  
\eqn 
and how fast $\f\in \Eg$ decays as $|x|\to \infty$  
depends on how many first moments of $V\f$ vanish:  
For $k=0,1, \dots$, 
\bqn \lbeq(asym-k)
\la V, x^{\alpha}\f\ra=0 \ \mbox{for 
$|\alpha|\leq k$} \ \Rightarrow \ 
|\f(x)| \leq C\ax^{-(2+k)}, \quad x \in \R^3,
\eqn  
whereas for resonances  
\bqn \lbeq(reso-asymp)
\f(x) = c|x|^{-1} + O(|x|^{-2}), 
\quad c=- \la V, \f \ra/4\pi \not=0.
\eqn
We shall often use Schur's lemma that the integral operator 
\bqn \lbeq(int-op)
Ku(x)= \int_Y K(x,y) d\n(y)
\eqn 
is bounded from $L^p(Y,d\n)$ to $L^p(X,d\m)$ for all 
$1\leq p \leq \infty$ if $K(x,y)$ satisfies  
\bqn \lbeq(bbb)
\sup_y \int_{X}|K(x,y)|d\m(x)< \infty, \quad   
\sup_x \int_{Y}|K(x,y)|d\n(y) <\infty.  
\eqn 
We say that the integral kernel $K(x,y)$ is  
{\it admissible} if it satisfies \refeq(bbb). 
We often identify the integral operator $K$ defined 
by \refeq(int-op) with its integral kernel $K(x,y)$ 
and say that $K(x,y)$ is an $L^p$-bounded kernel if $K$ 
is bounded in $L^p$. 
We write $\chi(F)$ for 
the characteristic function of the set $F$ and $a\absleq b$ 
means $|a|\leq |b|$. 

\paragraph{Acknowledgement} This work has been completed while 
the author is on leave from Gakushuin University and is 
visiting SISSA in Trieste. He would like 
to express his sincere gratitude to both institutions 
for making such a visit possible and for the warm 
hospitality in SISSA, to Gianfausto Dell'Antonio 
and Alessandro Michelangeli in particular.

\section{Reduction to the low energy analysis} 

We prove the theorem only for $W_{-}$ and write $W_{-}=W$ 
in the sequel. The conjugation ${\Cg}u(x)=\overline{u(x)}$ 
changes the direction of time and results for $W_{+}$ follows 
automatically from the ones    
for $W_{-}$. We write for $\lam \in \C^{+}$, $\C^{+}$ 
being the upper half plane,  
\[
G_0(\lam)= (H_0-\lam^2)^{-1}, \quad 
G(\lam)= (H-\lam^2)^{-1}.
\]
The limiting absorption principle and the absence of positive 
eigenvalues imply that, for $\s>1$, boundary values 
of $G_0(\lam)$ for $\lam\in \R$ and 
$G(\lam)$ for $\lam \in \R \setminus \{0\}$ exist in 
$\Bb(L^2_{\s}, L^2_{-\s})$ and $W$ can be represented via 
the boundary values in the following form 
(e.g. \cite{Ag, Ku-0}):  
\bqn 
Wu = u -\frac{1}{\pi i}\int^\infty_{0}  
G(\lam)V(G_0(\lam)-G_0(-\lam))u\lam d\lam 
\lbeq(stationary)
\eqn  

We decompose $W$ into the high and the low energy parts 
\bqn \lbeq(lh-decom)
W=W_{>} + W_{<} \equiv W \Psi(H_0) + 
W \Phi(H_0),
\eqn 
by using the cut off functions $\Phi\in C_0^\infty(\R)$ 
and $\Psi\in C^\infty(\R)$ such that 
\[
\Phi(\lam^2) + \Psi(\lam^2) \equiv 1, 
\quad \Phi(\lam^2)=1 \ \mbox{near $\lam=0$  and 
$\Phi(\lam^2)=0$ for $|\lam|>\lam_0$}
\] 
for a small constant $\lam_0>0$. 
We have proven in the previous paper \cite{Y-odd}  
that, under the assumption of this paper, 
$W_{>}$ is bounded in $L^p(\R^3)$ for all 
$1\leq p \leq \infty$ and we have nothing 
to add in this paper for $W_{>}$. Thus, 
in what follows, we shall be devoted to studying 
the low energy part: 
\bqn 
W_{<} = \Phi(H_0)  -
\frac{1}{\pi i}\int^\infty_{0}  
G(\lam)V(G_0(\lam)-G_0(-\lam))\lam \Phi(H_0) d\lam. 
\lbeq(stationary-l)
\eqn 
Evidently $\Phi(H_0)\in \Bb(L^p(\R^3))$ for all 
$1\leq p \leq \infty$ and we have only to study 
the operator $Z$ defined by the integral of 
\refeq(stationary-l), which we rewrite as  
\bqn \lbeq(z)
Z u = -\frac{1}{\pi i}\int^\infty_{0}  
G_0(\lam)V(1+ G_0(\lam)V)^{-1}
(G_0(\lam)-G_0(-\lam))\lam F(\lam)u d\lam 
\eqn 
by using the resolvent identity  
$G(\lam)V= G_0(\lam)V(1+ G_0(\lam)V)^{-1}$ for $\lam>0$   
and by defining $F(\lam)= \Phi(\lam^2)$.

\subsection{Low energy behavior of $(1+ G_0(\lam)V)^{-1}$. }

Following \cite{JK}, we say that $H$ is of exceptional type 
of {\it the first kind} if $\Eg=\{0\}$, 
{\it the second} if $\Eg=\Ng$ and {\it the third kind} if 
$\{0\}\subsetneq \Eg \subsetneq \Ng$. The orthogonal 
projection in $\Hg$ onto the eigenspace $\Eg$ will be 
denoted by $P$. We let $D_0, D_1, \dots$ be the 
integral 
operators defined by 
\[
D_j u(x) = \frac1{4{\pi}j!} \int_{\R^3}|x-y|^{j-1}u(y) dy, 
\quad j=0,1, \dots.
\]
so that we have a formal Taylor expansion  
\[
G_0(\lam)u(x)= \frac1{4\pi}\int_{\R^3}
\frac{e^{i\lam|x-y|}}{|x-y|}u(y)dy 
=\sum_{j=1}^\infty (i\lam)^j D_j u .
\]
If $H$ is of exceptional type of the third kind, $-(Vu, u)$ is an 
inner product of $\Ng$ and there exists a unique $\p\in \Ng$ 
such that 
\[
-(V\p, u)=0, \quad \forall u \in \Eg, \quad 
-(V\p,\p)=1 \ \mbox{and}\ (1, V\p) >0. 
\]
We define 
\bqn \lbeq(canonical)
\ph=\p + P VD_2 V \p\in \Ng
\eqn 
and call it {\it the canonical resonance} (\cite{JK}). 
If $H$ is of exceptional type of the first kind, 
then $\dim\Ng=1$ and there is a 
unique $\ph\in \Ng$ such that $-(V\ph, \ph)=1$ 
and $(1, V\ph) >0$ and we call this the canonical resonance. 
We have the following result (see e.g. \cite{Y-odd}). 

\bgprp \lbprp(3dim) Let $V$ satisfy $|V(x)|\leq C\ax^{-\delta}$ 
for some $\delta>3$. Suppose that $H$ is of exceptional 
type of the third kind and let $\ph$ be the canonical 
resonance and $a = 4\pi i |\la V,\ph \ra|^{-2}$. Then:  
\bqn \lbeq(res-1)
(I+G_0(\lam)V)^{-1}
=\frac{PV}{\lam^2} - \frac{PV D_3VPV}{\lam}  
- \frac{a}{\lam}|\varphi\ra \la \varphi| V  + E(\lam), 
\eqn 
where $E(\lam)$ is the operator valued function which,  
when substituted for $(1+ G_0(\lam)V)^{-1}$ in \refeq(z),  
produces the operator which is bounded in $L^p$ 
for all $1\leq p \leq \infty$. If $H$ is of exceptional 
type of the first or the second kind, \refeq(res-1) still 
holds with $P=0$ or $\ph=0$ respectively. 
\edprp

\section{$L^1$-unboundedness with resonances}

If zero energy resonances are present, then \refprp(3dim) shows 
that their contribution to the wave operator 
$W$ is given via the canonical resonance $\ph$ by 
\bqn \lbeq(f-k1)
Z_{r}u= -\frac{ia}{\pi}\int^\infty_{0}  
G_0(\lam)V\ph\ra 
\la V\ph |(G_0(\lam)-G_0(-\lam))u \ra F(\lam) d\lam, 
\eqn
where $a= 4\pi i|\la V, \ph\ra|^{-2}\not=0$. 
We show that $Z_{r}$ is bounded in $L^p$ for $1<p<3$ but not for 
$p =1$. 
\bglm \lblm(Schur)
Suppose that 
$K(x,y)\absleq C \ax^{-1}\ay^{-1}\la |x|-|y|\ra^{-2-\ep}$ 
for some $\ep\geq 0$. Then, 
\[
Ku(x) = \int_{\R^3} K(x,y) u(y) dy 
\]
is bounded in $L^p$ for any $1\leq p \leq \infty$ 
if $\ep>0$ and, for $1<p<\infty$ if $\ep=0$. 
\edlm 
\bgpf For $\ep>0$, the lemma follows from 
Schur's lemma. When $\ep=0$, we have 
\[
Ku(x)\absleq C\int_{0}^\infty 
\frac{r^{2-\frac{2}{p}} r^{\frac{2}{p}} |M_u(r)| dr}
{\ax \la r \ra \la |x|-r\ra^2}, \quad 
M_u(r)= \int_{{\mathbb S}^2}u(r\w)d\w . 
\]
Since the right side is rotationally invariant, we have 
\begin{align*}
\|Ku\|_p^p & \leq C 
\int_0^\infty  
\left(
\int_{0}^\infty 
\frac{\rho^{\frac{2}{p}} r^{2-\frac{2}{p}} 
(r^{\frac{2}{p}} |M_u(r)|) dr}
{\la \rho \ra \la r \ra \la \rho -r\ra^2} 
\right)^p d\rho \\ 
& \leq C 
\int_0^\infty 
\left(
\int_{0}^\infty 
\frac{\la \rho \ra^{\frac{2}{p}-1} 
\la r \ra ^{1-\frac{2}{p}} 
(r^{\frac{2}{p}} |M_u(r)|) dr}
{\la \rho-r\ra^2} 
\right)^p d\rho
\end{align*}
We may estimate 
$\la \rho \ra^{\frac{2}{p}-1} \la r \ra ^{1-\frac{2}{p}}$ 
by 
$C\la \rho-r \ra^{\frac{2}{p}-1}$ if $p\leq 2$ 
and 
$C\la \rho-r \ra^{1-\frac{2}{p}}$ if $p\geq 2$. 
It follows that unless $p=1$ or $p=\infty$ we have with 
$\gamma>1$ that 
\[
\|Ku\|_p \leq 
\left(
\int_0^\infty  
\left(
\int_0^\infty 
\frac{
r^{\frac{2}{p}}
|M_u(r)|dr
}
{\la \rho-r\ra^\gamma} 
\right)^p 
d\rho
\right)^{1/p}
\]
and Young's and H\"older's inequalities imply 
\[
\|Ku\|_p 
\leq C \left(\int_0^\infty  r^2 |M_u(r)|^p dr \right)^{1/p}
\leq C \|u\|_p .
\]
This completes the proof of lemma. 
\edpf 

\bgprp 
\lbprp(1st) Let $Z_r$ be the operator defined by \refeq(f-k1). 
Then, 
$Z_r$ is bounded in $L^p$ for $1<p<3$ but not for $p=1$. 
\edprp 
\bgpf It is known that $Z_r$ is bounded in $L^p$ for 
$1<p<3$ (\cite{Y-doc}). Nevertheless, we give the proof  
for $1<p<3$ which is different from the one given in \cite{Y-doc}. 
The integral kernel of $Z_r$ is given by 
\bqn \lbeq(zs-ker)
Z_r(x,y)= \sum_{\pm} \frac{\mp ia}{\pi}\int_0^\infty \int 
e^{i\lam (|x-z| \pm |w-y|)}\frac{(V\f)(z)(V\f)(w)F(\lam)}
{16\pi^2|x-z||w-y|}dw dzd\lam. 
\eqn 
Since $F\in C_0^\infty([0,\infty))$, we immediately see that,  
with a constant $C>0$,  
\bqn
Z_r(x,y)\absleq C \int \frac{|(V\f)(z)(V\f)(w)|}
{|x-z||w-y|}dw dz \leq \frac{C}{\ax\ay}
\eqn 
and $\chi(||x|-|y||\leq 1)Z_r(x,y)$ and 
$\chi(||x|^2-|y|^2|\leq 1)Z_r(x,y)$ are admissible kernels. 
Indeed, we have   
\[
\sup_y \int_{||x|-|y||\leq 1} \frac{dx}{\ax\ay} 
=\sup_x \int_{||x|-|y||\leq 1} \frac{dy}{\ax\ay}
= C <\infty, 
\]
$\{(x,y)\colon ||x|^2-|y|^2|\leq 1, ||x|-|y||> 1 \}
\subset \{(x,y)\colon |x|<1, |y|<1 \}$ and 
$C\ax^{-1}\ay^{-1}$ is obviously admissible on 
$\{(x,y)\colon |x|<1, |y|<1 \}$. 
Thus {\it we may and do ignore the parts of 
$Z_r(x,y)$ where $||x|-|y||<1$ or $||x|^2-|y|^2|<1$ in 
the proof}. 
We decompose the exponential functions 
$e^{i\lam |x-z|}$ and $e^{i\lam |w-y|}$ in the form 
\begin{gather} \lbeq(edec-1)
e^{i\lam |x-z|}= e^{i\lam |x|}+ e^{i\lam |x|}r(\lam,x,z),\quad  
r(\lam,x,z)= e^{i\lam (|x-z|-|x|)}-1,  \\
e^{i\lam |w-y|}= e^{i\lam |y|}+ e^{i\lam |y|}r(\lam,y,w),\quad  
r(\lam,y,w)= e^{i\lam (|w-y|-|y|)}-1,  
\lbeq(edec-2)
\end{gather}
and write $Z_r(x,y)$ as a sum of four kernels:
\[
Z_r(x,y) = \frac{-ia}{\pi} \sum_{j=1}^4  
\iint_{\R^6}\frac{(V\f)(z)(V\f)(w)}{16\pi^2|x-z||w-y|}
F_j(x,z,w,z)dw dz 
\]
where $F_j$, $j=1,2,3,4$ are respectively given by   
\begin{gather*} 
F_1  = F_1(x,z,w,y)=\sum \pm \int_0^\infty
e^{i\lam (|x|\pm |y|)}r(\lam, x, z)r(\pm \lam, y, w)F(\lam)
d\lam, \\ 
F_2 = F_2(x,w,y) = \sum \pm 
\int_0^\infty e^{i\lam (|x|\pm |y|)} r(\pm \lam, y,w)
F(\lam) d\lam, \\ 
F_3 = F_3(x,z,y)= \sum \pm \int_0^\infty 
e^{i\lam (|x|\pm |y|)} r(\lam, x, z)F(\lam)d\lam, \\ 
F_4= F_4 (x,y) = \sum \pm \int_0^\infty 
e^{i\lam (|x|\pm |y|)} F(\lam)d\lam.
\end{gather*}
Here and hereafter the sum $\sum$ is taken for $+$ and $-$.
We estimate $F_1, \dots, F_4$ using integration by parts. 
We use the following properties of $r(t,x,y)$ for $k=1,2, 3$:   
\begin{gather} \lbeq(r-prop)
r(0,x,y)=0, \quad  
\pa_{\lam}(r(\pm \lam,x,y))\vert_{\lam=0}= \pm i(|x-y|-|x|), \\
\quad \pa_\lam^k r(\lam,x,y)\leq |y|^k. \lbeq(r-k)
\end{gather}

\noindent 
(1) We first show that 
$Z_1(x,y)$ is an admissible kernel. 
We apply integration by parts three times to 
$F_1$. Then, \refeq(r-prop) and \refeq(r-k) imply  
\begin{align*}
& F_1(x,z,w,y)= \frac{\mp i}{(|x|\pm |y|)^3}
\pa_{\lam}^2 
\{
r(\lam, x, z)r(\pm \lam, y,w)F(\lam)
\}\vert_{\lam=0}  
\\
& \quad + \frac{\mp i}{(|x|\pm |y|)^3}
\int_0^\infty e^{i\lam (|x|\pm |y|)} \pa_{\lam}^3 \{
r(\lam, x, z)r(\pm \lam, y, w)F(\lam)\} d\lam   \\ 
& \quad \absleq \sum C \frac{(1+|z|+|w|)^3}{(|x|\pm |y|)^3} 
\leq C \frac{(1+|z|+|w|)^3}{(|x|-|y|)^3} .
\end{align*}
It follows that 
\[
Z_{1}(x,y) \absleq C \int_{\R^6} 
\frac{(1+|z|+|w|)^3|(V\f)(z)(V\f)(w)|}
{(|x|-|y|)^3 |x-z||w-y|}dw dz 
\leq  \frac{C}{(|x|-|y|)^3 \ax\ay}
\]
and  $Z_1(x,y)$ is admissible by virtue of \reflm(Schur) 
(recall that we are ignoring $(x,y)$ with  
$||x|-|y||<1$ or $||x|^2-|y|^2|<1$). 

\noindent 
(2) We apply integration parts twice to $F_2$ and write 
it in the form 
\[
\sum \frac{-i(|w-y|-|y|)}{(|x|\pm |y|)^2} 
+ \sum \mp 
\int_0^\infty 
e^{i\lam (|x|\pm |y|)}
\frac{\pa_{\lam}^2 \{r(\pm \lam, y, w) F(\lam)\}}
{(|x|\pm |y|)^2}d\lam.  
\]
After another integration by parts we see that the integral terms 
are bounded by $C(1+|w|)^3 (|x|\pm |y|)^{-3}$ and, 
when inserted into $Z_2(x,y)$, they produce admissible kernels 
bounded by $C\ax^{-1}\ay^{-1}(|x|\pm |y|)^{-3}$.  
Thus, modulo the admissible kernel 
\bqn 
Z_2(x,y)\equiv \sum \frac{a}{\pi}\f(x) 
\int_{\R^3} \frac{(|w-y|-|y|)(V\f)(w)}
{4\pi(|x|\pm |y|)^2\cdot |w-y|}dw.   \lbeq(z2-mod)
\eqn 
Note that this is bounded in modulus by 
$C\ax^{-1}\ay^{-1}(|x|-|y|)^{-2}$ and $Z_2$ is  bounded 
in $L^p$ for any $1<p<\infty$ by virtue of \reflm(Schur). 

\noindent 
(3) For $F_3$, we apply integration by twice as in (2): 
\[
F_3= \sum \mp 
\frac{i(|z-x|-|x|)}{(|x|\pm |y|)^2}
+ \sum_{\pm}\mp \int_0^\infty 
e^{i\lam (|x|\pm |y|)}
\frac{\pa_{\lam}^2 \{r(\lam, x, z) F(\lam)\}}
{(|x|\pm |y|)^2}d\lam .
\]
By applying integration by parts once more as in (2) 
we see  that the integral terms are bounded by 
$C(1+|z|)^3 (|x|\pm |y|)^{-3}$ and their sum produces 
the  kernel bounded by $C(|x|-|y|)^{-3}\ax^{-1}\ay^{-1}$
when inserted into $Z_3(x,y)$, which is admissible. Thus 
modulo the admissible 
kernel 
\begin{align*}
& Z_3(x,y) \equiv \sum \frac{\mp a}{\pi}
\int_{\R^6}
\frac{(|z-x|-|x|)(V\f)(z)(V\f)(w)}
{16\pi^2(|x|\pm|y|)^2 |x-z||w-y|}dzdw \\
& \absleq \frac{4a|x||y|\f(y)}{\pi (|x|+|y|)^2(|x|-|y|)^2}
\int_{\R^3}\frac{|z||(V\f)(z)|}{4\pi |x-z|}dz
\absleq \frac{C}{(|x|+|y|)^2(|x|-|y|)^2}, 
\end{align*}
where we used \refeq(fng) and \refeq(asym-k) for $k=0$ 
in the second and the third step respectively. 
Thus, $Z_3(x,y)$ is admissible. 

\noindent 
(4) Again an integration by parts shows that 
\[
F_4(x,y)= \sum \frac{\pm i}{(|x|\pm |y|)} 
+ \sum \frac{\pm i}{(|x|\pm |y|)}
\int_0^\infty e^{i\lam(|x|\pm |y|)}
F'(\lam)d\lam 
\]
Here $F'\in C_0^\infty((0,\infty))$ and the integral 
terms are bounded by $C \la |x|\pm |y| \ra^{-N}$ for any 
$N$. It follows that the sum of the integral terms produces 
an admissible kernel $C\la |x|-|y| \ra^{N} \ax^{-1} \ay^{-1}$ 
and, by virtue  of \refeq(fng), modulo the admissible kernel 
\[
Z_4(x,y) \equiv \sum \frac{\pm a\f(x)\f(y)}{\pi (|x|\pm |y|)}.
\] 

\noindent 
(5) We prove that $Z_r$ is unbounded in $L^1$. 
The combination of (1) to (4) implies that modulo 
admissible kernel $Z_r(x,y)$ is equal to 
\begin{multline} \lbeq(zred)
Z_{red}(x,y)=\sum \frac{a}{\pi}\f(x) \left(\frac1{(|x|+|y|)^2} 
+\frac1{(|x|-|y|)^2}\right)
\left(-c+ |y|\f(y)\right)  \\
+ \frac{a}{\pi}\left(
\frac{\f(x)\f(y)}{|x|+|y|}-\frac{\f(x)\f(y)}{|x|-|y|}\right) 
\equiv R_1(x,y) + R_2(x,y)
\end{multline}
where the constant $c\not=0$ is defined in \refeq(reso-asymp). 
We prove that the operator $Z_{red}$ defined by 
$Z_{red}(x,y)$ is unbounded in $L^1$ by contradiction. 
So assume that $Z_{red}$ is bounded in $L^1$. 
Take $u\in C_0^\infty(\R^3)$ such that 
$u(x)\geq 0$, $u(x)=0$ 
for $|x|\geq 1$ and $\int_{\R^3}u(x) dx=1$ and, define 
\[
u_n(x) = n^{3} u(nx), \quad 
f_n(x)= \int_{\R^3} Z_{red}(x,y) u_n(y) dy, \quad  
n=1,2, \dots. 
\] 
We have $\|u_n\|_1=1$, $n=1,2, \dots$. For any 
$2\leq |x|$, we evidently have  
\[
\lim_{n\to \infty} f_n(x)
=-\frac{2ac}{\pi}\frac{\f(x)}{|x|^2}.
\]
It follows by Fatou's lemma that 
\begin{multline}
\frac{2|ac|}{\pi}\int_{2<|x|}
\frac{|\f(x)|}{|x|^2}dx 
\leq \lim\inf_{n\to \infty}\int_{2<|x|}  |f_n(x)| dx \\
\leq \lim\inf_{n\to \infty} \|Z_{red} u_n\|_{L^1} 
\leq \|Z_{red}\|_{\Bb(L^1)}<\infty. \lbeq(contra)
\end{multline}
By virtue of \refeq(reso-asymp), $|\f(x)| \geq C |x|^{-1}$ 
for a constant $C>0$ for sufficiently large 
$|x|$ and \refeq(contra) cannot happen. Thus, $Z_{red}$ 
cannot be bounded in $L^1$. 

\noindent 
(6) We finally prove that $Z_{red}$ is bounded in $L^p$ 
for $1<p<3$. We have shown in (2) that $R_1(x,y)$ of the 
right of \refeq(zred) which is 
equal to \refeq(z2-mod) is $L^p$-bounded kernel for 
$1<p<\infty$ and it suffices to show that  
\[
R_2(x,y)= -\frac{2a}{\pi}\frac{\f(x)\f(y)|y|}{|x|^2-|y|^2}
\]
is an $L^p$-bounded kernel on $\{(x,y) \colon |x|^2- |y|^2\geq 1\}$ 
for $1<p<3$. Since $\f(x)|x| \in L^\infty$, 
it suffices to consider   
\[
Tu(x)=\int_{||x|^2-|y|^2|\geq 1} 
\frac{|x|^{-1}}{|x|^2-|y|^2}u(y) dy 
\]
Since $Tu(x)$ is spherically symmetric, we have by using polar 
coordinates and by changing variables 
\[
\|Tu\|_p^p \leq 2^{-1-p}\pi \int_0^\infty 
\rho^{\frac12-\frac{p}{2}}
\left(\int_{|\rho-r|\geq 1} 
\frac{r^{\frac12}}{\rho-r}|M_u(\sqrt{r})| 
dr\right)^p d\rho ,
\]
where $M_u(r)= \int_{{\mathbb S}^2}u(r\w)d\w$. 
Here $-1<\frac12-\frac{p}{2}<p-1$ if $1<p<3$ and 
$\rho^{\frac12-\frac{p}{2}}$ is an $(A)_p$ weight on 
$\R$. It follows by the weighted inequality 
(see e.g. Theorem 9.4.6 of \cite{Gr}) that 
\[
\|Tu\|_p^p 
\leq C \int_0^\infty r^{\frac12}|M_u(\sqrt{r})|^p dr 
\leq C \int_0^\infty r^2|M_u(r)|^p dr 
\leq C \|u\|_p . 
\]
This completes the proof. 
\edpf 

\bgrm \lbrm(res-b) If $\f \in \Eg$, then $|\f(x)|\leq  C \ax^{-2}$  
and  
\[
|R_1(x,y)|+ |R_2(x,y)| \leq 
C \ax^{-2}\ay^{-1}(|x|-|y|)^{-2}.
\]
Hence \reflm(Schur) implies both $R_1$ and $R_2$ 
are $L^p$-bounded kernels for all 
$1<p<\infty$. They are bounded also in $L^1$ 
because    
\[
\sup_{y}
\int_{||x|-|y||\geq 1} \frac{dx}{\ax^2 \ay(|x|-|y|)}
\leq 4\pi \sup_{y} 
\int_{0}^\infty \frac{dr}{\ay \la r-|y|\ra^2}
<\infty.
\]
\edrm  

\section{Contribution of zero-energy eigenfunctions}

The following proposition together with \refprp(3dim) and 
\refprp(1st) proves \refth(1). 

\bgprp \lbprp(2nd) Let $H$ be of exceptional type of 
the second kind. Then, $W_\pm$ is bounded in $L^p$ for 
$1\leq p<3$ in general. 
If $\la V, x^\alpha \f\ra=0$, $|\alpha|\leq 2$, for all 
$\f \in \Ng$, then $W_\pm$ is bounded in $L^p$ for all 
$1\leq p \leq \infty$. 
\edprp

If $H$ is of exceptional type of the second kind, then     
\bqn \lbeq(s2)
S(\lam)= \frac{P V}{\lam^2} -\frac{P V D_3 V P V}{\lam},
\eqn 
where $D_3$ is the operator of rank five with the  
kernel $-i|x-y|^2/4\pi$ and all $\f \in \Eg$ satisfy 
$|\f(x)| \leq C\ax^{-2}$ for a constant $C>0$. 
We take the real orthonormal basis $\{\f_1, \dots, \f_d\}$ 
of $\Eg$  and write $Z_{s}u= Z_{s0}u + Z_{s1}u$, where 
with $a_{jk}=i \pi^{-1}\la \f_j |VD_3V|\f_k\ra \in \R$,  
\bqn \lbeq(d3-zsem-0)
Z_{s0} u = \sum_{j,k=1}^d a_{jk} \int^\infty_{0}  
G_0(\lam)V\f_j \ra 
\la V\f_k |(G_0(\lam)-G_0(-\lam))u \ra F(\lam) d\lam
\eqn 
and, with extra singularity $\lam^{-1}$, 
\bqn 
\lbeq(d3-zsem-1)
Z_{s1} u = \sum_{j=1}^d \frac{i}{\pi}\int^\infty_{0}  
G_0(\lam)V\f_j \ra 
\la V\f_j |(G_0(\lam)-G_0(-\lam))u \ra F(\lam) 
\frac{d\lam}{\lam}.
\eqn 
 
\bglm \lblm(1) 
{\rm (1)} The operator $Z_{s0}$ is bounded in $L^p$ for 
any $1\leq p<\infty$. 

\noindent 
{\rm (2)} If all $\f\in \Eg$ satisfy 
$\int_{\R^3} x_j V(x)\f(x)dx=0$ for all $j=1,2,3$ then,  
$Z_{s0}$ is bounded in $L^p$ for all $1\leq p \leq \infty$. 
\edlm 
\bgpf $Z_{s0}$ is the sum of $Z_{s0,ij}$, 
$1\leq i,j \leq d$,  
whose kernels have the same structure as $Z_r(x,y)$ of \refeq(f-k1) 
with only change of the constant $a$ by $i\pi {a}_{ij}$ 
and of the canonical resonance $\ph$ 
by eigenfunctions $\f_i$ and $\f_j\in \Eg$, which satisfy all 
properties of $\ph$ which are necessary for proving \refprp(1st). 
Then, if we proceed as in the proof of \refprp(1st), all integral 
kernels which appear there are admissible 
except $R_{1,ij}(x,y)$ and $R_{2,ij}(x,y)$ which 
correspond to $R_1(x,y)$ and $R_2(x,y)$ of \refeq(zred). If 
$\f_i,\f_j\in \Eg$, however, as remarked in \refrm(res-b), 
they are estimated as 
\bqn \lbeq(s0-re)
R_{1,ij}(x,y)=
\sum_{\pm} C_{ij}\frac{\f_i(x)|y|\f_j(y)}{(|x|\pm |y|)^2}
\absleq C \sum_{\pm} \frac{\ax^{-2}\ay^{-1}}{(|x| \pm |y|)^2}
\eqn 
and likewise for $R_{2,ij}(x,y)$. Thus, they 
are $L^p$-bounded for $1\leq p<\infty$. If 
$\f_i, \f_j\in \Eg$ further satisfy the 
condition of (2), the right of \refeq(s0-re) is bounded by 
$\ax^{-3}\ay^{-2}\la |x| -|y|\ra^{-2}$ and they are admissible 
kernels (recall that we are ignoring 
the part $\{(x,y) \colon ||x|-|y||\leq 1\}$). 
\edpf 

\bglm \lblm(3) 
{\rm (1)} The operator 
$Z_{s1}$ is bounded in $L^p$ for $1\leq p<3$. 

\noindent 
{\rm (2)} 
If all $\f \in \Eg$ satisfy 
$\la V, x^{\alpha}\f \ra=0$ for $|\alpha|\leq 2$. Then 
$Z_{s1}$ is bounded in $L^p$ for all $1\leq p \leq \infty$. 
\edlm 
\bgpf Define for $j=1, \dots, d$ that 
\bqn \lbeq(sec-1)
Z_{s1,j}u = \frac{i}{\pi}\int_0^\infty 
G_0(\lam)|V\f_j\ra 
\la V\f_j |(G_0(\lam) -G_0(-\lam))u\ra F(\lam)\lam^{-1}d\lam,  
\eqn 
so that $Z_{s1}u= \sum_{j=1}^d Z_{s1,j} u$. We prove the 
lemma only for $Z_{s1,1}$, which we denote by $Z$ for short,  
as the proof for others is similar.    
As in the proof of \refprp(1st), we decompose $Z$ as 
\[
Zu = Z_1u + Z_2 u + Z_3 u + Z_4 u 
\]
by splitting $e^{i\lam |x-y|}$ and etc. as 
$e^{i\lam |x-y|}= e^{i\lam |x|}+ e^{i\lam|x|}r(\lam,x,y)$ 
and etc. Thus the integral kernels of $Z_1, \dots, Z_4$ are given 
respectively by 
\begin{gather*} 
Z_1  = \sum_{\pm} \frac{\pm i}{\pi}\int_0^\infty \int 
e^{i\lam (|x|\pm |y|)} \frac{r(\lam, x, z)r(\pm \lam, w, y)
 V\f(z)V\f(w)F(\lam)}{16\pi^2|x-z||w-y|}dw dz\frac{d\lam}{\lam}, \\ 
Z_2 = \sum_{\pm}\frac{\mp i}{\pi}\int_0^\infty \int 
e^{i\lam (|x|\pm |y|)} \frac{r(\pm \lam, w, y)
 \f(x)(V\f)(w)F(\lam)}{4\pi |w-y|}dw   
\frac{d\lam}{\lam}, \\ 
Z_3 = \sum_{\pm}\frac{\mp i}{\pi}\int_0^\infty \int 
e^{i\lam (|x|\pm |y|)} \frac{r(\lam, x, z)
 (V\f)(z)\f(y)F(\lam)}{4\pi|x-z|} dz 
\frac{d\lam}{\lam}, \\ 
Z_4 u = \frac{i}{\pi}\int_0^\infty  
\sum_{\pm} \pm e^{i\lam (|x|\pm |y|)} \f(x)\f(y)F(\lam) 
\frac{d\lam}{\lam}, 
\end{gather*}
where we used \refeq(fng) for simplifying  $Z_2, Z_3$ and $Z_4$ 
and put the sign $\sum$ inside the integral as it diverges 
separately. We write 
\bqn 
r(\lam,x,y)= i\lam (|x-y|-|x|)r_1(\lam, x,y), \quad 
r_1= \int_0^1 e^{i\lam (|x-y|-|x|)\theta}d\theta  \lbeq(r-a)
\eqn 
and etc. We have for $k=0,1, \dots$, 
\bqn \lbeq(r-1)
r^{(k)}_1(\lam,x,y)\absleq |y|^k/(k+1) .  
\eqn 
We estimate $Z_1u, \dots, Z_4 u$ individually in the 
following four lemmas. We obviously have 
$Z_j(x,y)\absleq \ax^{-1}\ay^{-1}$  for $j=1,2,3$ and 
this holds also for $j=4$ as 
$|\f(y)(e^{i\lam|y|}-e^{-i\lam|y|})|\lam^{-1}\leq |y||\f(y)| 
\leq C\ay^{-1}$. Thus, their parts on 
\bqn 
\lbeq(ignore)
\{(x,y)\colon ||x|-|y||<1\}\cup  
\{(x,y)\colon ||x|^2-|y|^2|<1\} \cup  \{(x,y)\colon |x|+|y|<1\} 
\eqn 
are admissible kernels and we again ignore this part in the 
following consideration. 

\bglm \lblm(Z-1) Modulo an admissible kernel we have 
\[
Z_{1}(x,y)\equiv  \sum_{\pm} 
\frac{i}{\pi}\frac{|x| |y|\f(x)\f(y)}
{(|x|\pm |y|)^2}
\]
and $Z_1$ is an $L^p$-bounded kernel for $1<p<\infty$. 
If $\f\in \Eg$ satisfies 
$\la V, x^\alpha \f\ra=0$ for $|\alpha|\leq 1$, $Z_1(x,y)$ is  
admissible. 
\edlm 
\bgpf By virtue of \refeq(r-a), we have 
\begin{align*} 
& Z_1(x,y) =  -\frac{i}{\pi} \int 
\left(\sum_{\pm} \int_0^\infty 
e^{i\lam (|x|\pm |y|)}r_1(\lam, x, z)r_1(\pm \lam, w, y)  
\lam F(\lam) d\lam\right) \\
& \qquad \qquad \times \frac{(|x-z|-|x|)(|w-y|-|y|)
 (V\f)(z)(V\f)(w)}{16\pi^2
|x-z||w-y|}dw dz \\
& = -\frac{i}{\pi} \int W_1(x,z,w, y)
\frac{(|x-z|-|x|)(|w-y|-|y|)
 (V\f)(z)(V\f)(w)}{16\pi^2|x-z||w-y|}dw dz,
\end{align*}
where the definition of $W_1$ should be obvious. 
We apply integration by parts twice to $W_1(x,z,w, y)$ 
and obtain    
\begin{gather} \lbeq(w-1)
W_1(x,z,w,y)= \sum_{\pm} \frac{-1}{(|x|\pm |y|)^2}
+ Y_1(x,z,w,y), \\
Y_1= \sum_{\pm} \frac{-1}{(|x|\pm |y|)^2} \
\int_0^\infty 
e^{i\lam (|x|\pm |y|)}
(r_1(\lam, x, z)r_1(\pm \lam, w, y)  \lam F(\lam))''  d\lam. 
\notag
\end{gather} 
By using $\int (V\f)(x)dx=0$ and \refeq(fng), we have 
\bqn \lbeq(ef)
\int \frac{(|x-z|-|x|)(V\f)(z)}{4\pi|x-z|}dz
= |x|\f(x)
\eqn 
and the like for the integral involving $(V\f)(w)$. 
Thus, the contribution to $Z_1(x,y)$ of 
the boundary term in 
\refeq(w-1) is given by  
\bqn \lbeq(wb)
\frac{i}{\pi}\sum_{\pm} \frac{|x| |y|\f(x)\f(y)}{(|x|\pm |y|)^2}.
\eqn 
By virtue of \reflm(Schur) both $+$ and $-$ terms of \refeq(wb) are 
$L^p$ bounded kernels for all $1<p<\infty$ and they are admissible 
if $\int_{\R^3}x_j (V\f)(x)dx=0$ for $j=1,2,3$.  
Further integration by parts and \refeq(r-1) imply    
\[
Y_1(x,z,w,y) 
\absleq C \sum_{\pm} \frac{(1+ |z|+|w|)^3}{(|x|\pm |y|)^3}.
\]
Thus, the contribution of $Y_1$ to 
$Z_1(x,y)$ produces the kernel which is bounded in modulus by 
\begin{align*}
C \sum_{\pm} & \int 
\frac{|z||w|(1+ |z|+|w|)^3 |(V\f)(z)(V\f)(w)|}
{|x-z||w-y|(|x|\pm |y|)^3}dw dz  \\
& \leq \sum_{\pm}  \frac{C}{\ax \ay (|x|\pm |y|)^3}
\end{align*}
and, hence, is admissible. This proves the lemma. 
\edpf

\bglm \lblm(Z-2) Modulo an admissible kernel  
\bqn \lbeq(z-2)
Z_{2}(x,y)\equiv 
\frac{2i}{\pi}\frac{|x| \f(x) |y|\f(y)}
{|x|^2\pm |y|^2} 
\eqn 
and $Z_2$ is bounded for all $1<p<\infty$. If $\f\in \Eg$ satisfies 
$\int_{\R^3}x_j (V\f)(x)dx=0$ for $j=1,2,3$, in addition, 
then $Z_2(x,y)$ is admissible. 
\edlm 
\bgpf The proof goes in parallel with that of \reflm(Z-1). 
Using \refeq(r-a), we write $Z_2(x,y)$ 
in the form   
\[ 
\frac{1}{\pi}\int \left(\sum_{\pm} \int_0^\infty 
e^{i\lam (|x|\pm |y|)}r_1(\pm \lam, w, y) F(\lam) d\lam\right)
\frac{(|w-y|-|y|)\f(x)(V\f)(w)}{4\pi |w-y|}dw.
\]
Write $W_2=W_2(x,w,y)$ for the function inside the parenthesis. 
By integration by parts, we have 
\[
W_2
= \sum_{\pm} 
\frac{i}{|x|\pm |y|}+ \sum_{\pm} 
\frac{i}{|x|\pm |y|} \int_0^\infty 
e^{i\lam (|x|\pm |y|)}(r_1(\pm \lam, w, y)F(\lam))' d\lam .
\]
The contribution to $Z_2(x,y)$ of the boundary term for $W_2$ 
is given by 
\[ 
\sum_{\pm} \frac{i}{\pi} \int 
\frac{(|w-y|-|y|)\f(x)(V\f)(w)}{(|x|\pm |y|)\cdot 4\pi |w-y|}dw
=\sum_{\pm} \frac{i}{\pi}\frac{\f(x) |y|\f(y)}{|x|\pm |y|},
\]
which is equal to \refeq(z-2). 
Further integration by parts shows the integral term for $W_2$ 
is given by      
\[
Y_2(x,w,y) 
= \sum_{\mp i } \frac{\mp (|w-y|-|y|)}{2(|x|\pm |y|)^2}
+ O\left(\frac{C (1+|w|)^3 }{(|x|\pm |y|)^3}\right).
\]
The second term on the right produces in $Z_2(x,y)$ the admissible 
kernel bounded outside \refeq(ignore) by 
\[
C \frac{|\f(x)|\ay^{-1}}{\la |x|-|y|\ra^3} 
\]
and, the boundary term in $Y_2$ does in $Z_2(x,y)$ the kernel      
\[
\sum_{\pm i}\frac{\mp 1}{2(|x|\pm |y|)^2}\int_{\R^3} 
\frac{\f(x)(|w-y|-|y|)^2 (V\f)(w)}{4\pi^2|w-y|}dw
\absleq 
\frac{C|\f(x)||x||y|}{\ay (|x|^2 -|y|^2)^2 }, 
\]
which is again admissible outside \refeq(ignore). 
If $\f$ satisfies $\la V, x_j \f \ra=0$ for $j=1,2,3$, 
then $|\f(x)| \leq C\ax^{-3}$ and \refeq(z-2) also becomes 
admissible. This proves the lemma. 
\edpf 

\bglm \lblm(Z-3)
Modulo an admissible kernel  
\bqn \lbeq(z-3)
Z_{3}(x,y)\equiv \frac{-2i}{\pi} 
\frac{|x|\f(x)|y|\f(y)}{|x|^2-|y|^2}
\eqn 
and $Z_3$ is bounded for all $1<p<\infty$. 
If $\f\in \Eg$ satisfies 
$\int_{\R^3}x_j (V\f)(x)dx=0$ for $j=1,2,3$ in addition, 
then $Z_2(x,y)$ is an admissible kernel. 
\edlm 
\bgpf The proof goes in parallel with that of \reflm(Z-2). 
By using \refeq(r-a) and \refeq(ef) once more we write 
$Z_3(x,y)$ in the form  
\begin{gather}
Z_3(x,y)= \frac{1}{\pi}\int_{\R^3}
W_3(x,z,y)\frac{(|x-z|-|x|)(V\f)(z)\f(y)}
{4\pi |x-z|} dz . \lbeq(add-b) \\
W_3(x,z,y)= 
\sum_{\pm} \pm \int_0^\infty 
e^{i\lam (|x|\pm |y|)}r_1(\lam, x, z)F(\lam) d\lam.
\end{gather}
Application of integration by parts shows that   
\begin{gather}
W_3(x,z,y)= \sum_{\pm} \frac{\pm i}{|x|\pm |y|} 
+ Y_3(x,z,y), \lbeq(w-3b) \\  
Y_3(x,z,y)= \sum_{\pm} \frac{\pm i}{|x|\pm |y|}\int_0^\infty  
e^{i\lam (|x|\pm |y|)}(r_1(\lam, x, z)F(\lam))' d\lam. 
\lbeq(y-3)
\end{gather}
The contribution to $Z_3(x,y)$ of the boundary term of 
$W_3$ in \refeq(w-3b) is given by virtue of \refeq(ef) by 
\bqn 
\sum_{\pm} \frac{\pm i}{\pi}\int_{\R^3}
\frac{(|x-z|-|x|)(V\f)(z)\f(y)}
{(|x|\pm |y|)4\pi |x-z|} dz 
= \sum_{\pm} \frac{\pm i}{\pi} \frac{|x|\f(x)\f(y)}{|x|\pm |y|}
\eqn 
Further integration by parts implies 
\bqn \lbeq(add-a)
Y_3(x,z,y)= 
\sum_{\pm} \frac{\mp {i}(|x-z|-|x|)}{2(|x|\pm |y|)^2}
+ O\left(\frac{(1+|z|)^3}{(|x|\pm |y|)^3 }\right)
\eqn 
and substituting $Y_3$ for $W_3$ in \refeq(add-b) produces 
an admissible kernel. Indeed, from the boundary term on the 
right of \refeq(add-a) we have 
\[
\sum \frac{\mp {i}}{2(|x|\pm |y|)^2} 
\int_{\R^3}\frac{(|x-z|-|x|)^2 (V\f)(z)\f(y)}{4\pi|x-z|}dz 
\absleq \frac{C|x||y||\f(y)|}{(|x|^2-|y|^2)^2 \ax}  
\]
which is admissible and, from the second term the kernel 
bounded by   
\[
C \frac{|\f(y)|\ax^{-1}}{\la |x|-|y|\ra^3},
\]
which is also admissible. If $\f\in \Eg$ satisfies 
$\int_{\R^3}x_j (V\f)(x)dx=0$ for $j=1,2,3$ in addition, 
it is obvious that \refeq(z-3) becomes an admissible 
kernel. This concludes the proof of the lemma. 
\edpf 

For proving the following \reflm(Z-4), we use the next lemma. 
\bglm \lblm(K) 
\ben 
\item[{\rm (1)}] 
Suppose that $K(x,y)$ satisfies 
\bqn \lbeq(k)
K(x,y) \absleq C \int_{-1}^1 
\ax^{-2}\ay^{-1}\la |x|-\theta |y|\ra^{-\delta}d\theta, 
\quad  x,y \in \R^3,  
\eqn 
for some constants $\delta>1$ and $C>0$. Then, $K$ 
is $L^p$-bounded kernel for $1\leq p <3$. 
\item[{\rm (2)}]  
If $K$ 
satisfies \refeq(k) with $\ax^{-2}\ay^{-\kappa}$, $\kappa>2$  
in place of $\ax^{-2}\ay^{-1}$, then $K$ is admissible. 
\een
\edlm 
\bgpf (1) It suffices show that $K$ is bounded in $L^1$ 
and $L^p$ for $2<p<3$ by virtue of the interpolation theorem. 
By using polar coordinates, we estimate 
\bqn \lbeq(1seo)
\int_{\R^3} |K(x,y)|dx 
\leq C \int_0^1 \left(\int_0^\infty 
\frac{dr}{\la r-\theta |y|\ra^{\delta}}\right) \ay^{-1} d\theta
\leq C \ay^{-1}
\eqn 
and $K$ is bounded in $L^1$. We next let $2<p<3$. Minkowski's 
inequality implies 
\bqn \lbeq(z4pre)
\left(\int_{\R^3}|K(x,y)|^p dx \right)^{1/p}
\leq C \ay^{-1}\int_0^1 \left(\int_0^\infty 
\frac{r^2\la r\ra^{-2p}}{\la r-\theta |y|\ra^{p\delta}} 
dr\right)^{1/p}d\theta.
\eqn 
We denote the integrand $(\cdots)^{1/p}$ with respect to $\theta$ 
by $G(\theta |y|)$. 
It is obvious that $G(\theta|y|)\leq C$ if $\theta |y|\leq 2$, 
hence $\refeq(z4pre) \leq C$ for $|y|\leq 2$. Let 
$\theta |y|\geq 2$ and $|y|\geq 2$. We split as 
$(0,\infty)=I_1 \cup I_2$, 
\[
I_1= \{r>0 \colon \theta |y|/2 <r<3\theta |y|/2\}\  \mbox{and} \ 
I_2= \{r>0 \colon |r-\theta |y|| \geq \theta |y|/2\}
\]
and estimate 
$r^2\la r\ra^{-2p} \leq C \la \theta |y| \ra^{2-2p}$ on $I_1$ 
and 
$\la r-\theta |y|\ra^{-p\delta} \leq C\la \theta |y|\ra^{-p\delta}$ 
on $I_2$. Then,  
\[
G(\theta|y|)\leq \left\{\left(\int_{I_1}+ \int_{I_2}\right) 
\frac{r^2\la r\ra^{-2p}}{\la r-\theta |y|\ra^{p\delta}} 
dr\right\}^{1/p}\leq C (\la \theta |y| \ra^{\frac{2}{p}-2}
+\la \theta |y| \ra^{-\delta}). 
\] 
Since $\delta>1$ and $2<p<3$, we obtain for $|y|\geq 2$ that 
\[
\int_{0}^1\chi(\theta |y|\geq 2) G(\theta |y|) d\theta
\leq C\left(\int_{2/|y|}^1 
(\la \theta |y| \ra^{\frac{2}{p}-2}+ 
\la \theta |y| \ra^{-\delta})d\theta \right) \leq 
\frac{C}{\ay}. 
\]
Thus, for $2<p<3$, we have by Minkowski's and H\"older 
inequalities that 
\[
\|K u\|_p 
\leq \int_{\R^3}
\left(\int_{\R^3}|K(x,y)|^p dx \right)^{1/p} |u(y)|dy 
\leq C \|\ay^{-2}\|_{p'} \|u\|_p \leq C \|u\|_p ,
\]
where the dual exponent $p'=p(p-1)^{-1}>3/2$. 

\noindent  
(2) We have 
$\ax^{-2}\la |x|-\theta |y|\ra^{-\delta}
\leq C \la \theta |y|\ra^{-\min(2,\delta)}$
and 
\[
\sup_{x\in \R^3}\int_{\R^3} 
|K(x,y)|dy \leq C \int_{\R^3} \ay^{-\kappa-1}dy <\infty.
\]
This with \refeq(1seo) implies $K$ is admissible. 
\edpf 

\bglm \lblm(Z-4) \ben 
\item[{\rm (1)}] 
Modulo an $L^p$-bounded kernel for $1\leq p <3$,  
\bqn \lbeq(z-40)
Z_4(x,y)\equiv - \frac{i}{\pi}\f(x)\f(y)|y| \cdot \int_{-1}^1  
\frac{\chi(||x|+\theta|y||>1)}
{|x|+\theta|y|} d\theta .
\eqn 
\item[{\rm (2)}] $Z_4$ is bounded in $L^p$ for $3/2<p<3$. 
\item[{\rm (3)}] 
If $\la V, x^\alpha \f \ra=0$ for $|\alpha|\leq 1$, then 
\refeq(z-40) holds modulo the kernel which is $L^p$-bounded  
for all $1\leq p \leq \infty$. 
\een
\edlm 
\bgpf We have   
\begin{align}
Z_4(x,y)& = \frac{i}{\pi}\f(x)\f(y)
\int_0^\infty e^{i\lam|x|}(e^{i\lam|y|}-e^{-i\lam |y|})
\frac{F(\lam)}{\lam}d\lam \notag \\ 
& = -\frac{1}{\pi}\f(x)\f(y)|y|\int_{-1}^1 \left( 
\int_0^\infty 
e^{i\lam(|x|+\theta|y|)}F(\lam) d\lam\right) d\theta.
\lbeq(z-4)
\end{align}
By inserting 
$1= \chi(||x|+\theta|y||\leq 1) + \chi(||x|+\theta|y||>1)$ in 
front of $d\theta$ of \refeq(z-4), we split $Z_4(x,y)$ into 
two parts: 
\[
Z_4(x,y)= Z_4^{\leq} (x,y)+ Z_4^{>}(x,y).
\]
We clearly have 
\[
Z_4^{\leq} (x,y) \absleq 
\frac{2}{\pi}\int_0^1 |\f(x)\f(y)||y|
\chi(||x|-\theta|y||\leq 1)
d\theta 
\]
and \reflm(K) implies $Z_4^{\leq}$ is bounded in $L^p$ 
for $1\leq p<3$ in general and, is admissible  
if $\f$ satisfies $|\f(x)|\leq C \ax^{-3}$. We next 
consider $Z_4^{\geq}(x,y)$. 
By integration by parts, we have that 
\bqn \lbeq(z-4a)
\int_0^\infty e^{i\lam(|x|+\theta|y|)}F(\lam) d\lam
= \frac{i}{|x|+\theta|y|} + 
{i}\int_0^\infty 
\frac{e^{i\lam(|x|+\theta|y|)}F'(\lam)}{|x|+\theta|y|} d\lam
\eqn
and, as $F'\in C_0^\infty((0,\infty))$, the integral term is 
bounded for any $N=1,2, \dots$ by 
$C\la |x|+\theta|y|\ra^{-N}$ when 
$||x|+\theta|y||\geq 1$. It follows that its contribution of 
to $Z_4^{>}(x,y)$ produces the kernel bounded in 
modulus by 
\[
C \int_{-1}^1 \frac{|\f(x)\f(y)||y|}
{\la |x|+\theta|y|\ra^{N}} d\theta
\]
and \reflm(K) implies it enjoys the same property of $Z_4^{\leq}$ 
stated above. 
Since the contribution of the boundary term in \refeq(z-4a) 
to  $Z_4^{>}(x,y)$ is given by \refeq(z-40), statements 
(1) and (3) of the lemma are proved. 

\noindent 
(2) Since $Z_4(x,y)\absleq C|y| \f(x)\f(y)$, it is obvious 
that $\chi(|y|<2)Z_4(x,y)$ produces bounded operator in 
$L^p$ for $3/2<p<3$ and  we may ignore the part  
$\{y\in \R^3 \colon |y|<2\}$ of $Z_4(x,y)$. 
We estimate 
\begin{multline}
\int_{-1}^1 \frac{\chi(||x|+\theta|y||>1)}
{|x|+\theta|y|} d\theta
=\frac{1}{|y|}
\int_{|x|-|y|}^{|x|+|y|} \frac{\chi(|\theta |>1)}
{\theta} d\theta \\
\leq \frac{1}{|y|}
\int_{1}^{|x|+|y|} \frac{d\theta}
{\theta} = 
\frac{\log(|x|+|y|)}{|y|} 
\leq \frac{\log(1+|x|)+\log(1+|y|)}{|y|}.
\end{multline}
Then we have 
$Z_4(x,y) \absleq C(\log \ax +\log\ay) \f(x)\f(y)$ 
and $\log\ax\f(x)\in L^p(\R^p)$  for any 
$3/2<p $. Statement (2) follows. 
\edpf 

\paragraph{Completion of the proof of \reflmb(3).} 
Combining the results of previous four lemmas and observing 
that the boundary terms in \refeq(z-2) and \refeq(z-3) cancel 
each other, we see that $Z(x,y)$ is modulo the kernel 
which is bounded in $L^p$ for $1\leq p<3$ 
equals to the sum $K_0(x,y)=K_{01}(x,y)+ K_{02}(x,y)$ where 
\bqn 
K_{01}(x,y)= 
\sum_{\pm} 
\frac{i}{\pi}\frac{\chi(||x|-|y||\geq 1)|x| |y|\f(x)\f(y)}
{(|x|\pm |y|)^2} \lbeq(k01) 
\eqn 
and 
\bqn 
K_{02}(x,y)= -\frac{i}{\pi}\f(x)\f(y)|y| \cdot \int_{-1}^1  
\frac{\chi(||x|+\theta|y||>1)}
{|x|+\theta|y|} 
d\theta.
\eqn 
Since \reflms(Z-1,Z-2) already prove that $K_0$ is bounded in 
$L^p$ for $3/2<p<3$, it suffices to show by virtue of 
interpolation (\cite{BL}) that 
$K_0$ is also bounded in $L^1$. 
We write $|x|=|x| \pm |y| \mp |y|$ in \refeq(k01). Then, 
\begin{align}
K_{01}(x,y)& = 
\sum_{\pm} 
\frac{i}{\pi}\frac{\chi(||x|-|y||\geq 1) |y|\f(x)\f(y)} 
{|x|\pm |y|} \lbeq(k02-a) \\
& - \frac{i}{\pi} 
\frac{4\chi(||x|-|y||\geq 1)|x| |y|^3 \f(x)\f(y)}
{(|x|+|y|)^2(|x|-|y|)^2}.  \lbeq(k02)
\end{align}
We denote \refeq(k02) by $K_a(x,y)$. It is then evident 
that 
\[
K_a(x,y)\absleq C 
\frac{\chi(||x|-|y||\geq 1)|\f(x)||y|^2 \f(y)}
{(|x|-|y|)^2}
\]
and it produces a bounded operator in $L^1$. 
Thus, for concluding the proof of 
\reflm(3), it suffices show that   
\[
K_e(x,y) = \frac{i}{\pi}\f(x)\f(y)|y|
\left(\sum_{\pm} \frac{\chi(||x|-|y||\geq 1)}{|x|\pm |y|} 
- \int_{-1}^1 \frac{\chi(||x|+\theta|y||>1)}
{|x|+\theta|y|} d\theta \right) 
\]
is an $L^1$-bounded integral kernel. We first remark that 
\begin{multline*}
\f(x)\f(y)|y|
\left(\frac{\chi(||x|-|y||\geq 1)}{|x|\pm |y|}
-\frac{1}{\la |x|\pm |y|\ra}\right) 
\\
= \frac{\f(x)\f(y)|y|\chi(||x|-|y||\geq 1)}
{(|x|\pm |y|)\la |x|\pm |y|\ra
\{(|x|\pm |y|)+\la |x|\pm |y|\ra\}} 
-\frac{\chi(||x| \pm |y||\leq 1)}{\la |x|\pm |y|\ra}
\end{multline*}
is an $L^1$ bounded integral kernel. 
We can likewise show that 
\[
\int_{-1}^1 \f(x)\f(y)|y|\left( 
\frac{\chi(||x|+\theta|y||>1)}
{|x|+\theta|y|} -\frac{1}{\la |x|+\theta |y|\ra}\right)
d\theta 
\]
is an $L^1$-bounded kernel. Finally, observing that 
\[
\frac1{\la a\ra} -\frac1{\la b\ra }
= \frac{\la b\ra -\la a\ra}{\la a\ra\la b\ra}
= \frac{b^2 -a^2}{\la a\ra\la b\ra(\la a\ra+\la b\ra)}
\absleq 
\frac{|b -a|}{\la a\ra\la b\ra},
\]
we estimate 
\begin{align} 
& \frac{i}{\pi}\f(x)\f(y)|y|
\left(\sum_{\pm} \frac{1}{\la |x|\pm |y| \ra} 
- \int_{-1}^1 \frac{d\theta}{\la |x|+\theta|y|\ra} \right) 
\lbeq(final) 
\\
& \qquad = \sum_{\pm} 
\frac{i}{\pi}\f(x)\f(y)|y| 
\int_0^1 \left(\frac{1}{\la |x|\pm |y| \ra}-
\frac{1}{\la |x|\pm \theta|y|\ra} \right)d\theta \\
& \qquad 
\absleq 
\sum_{\pm} 
\frac{i}{\pi}\f(x)\f(y)|y| 
\int_0^1 \frac{(1-\theta)|y|}
{\la |x|\pm |y| \ra \la |x|\pm \theta|y|\ra} 
d\theta .
\end{align}
Then, we have 
\begin{multline*}
\int_{\R^3} \frac{\f(x)\f(y)|y|^2}
{\la |x| \pm |y| \ra \la |x| \pm \theta|y|\ra}dx 
\leq C \int_0^\infty 
\frac{dr}{\la r \pm  |y| \ra \la r  \pm \theta|y|\ra}\\
\leq \frac{C}2 
\int_{-\infty}^\infty 
\left( \frac{1}{\la r \pm |y| \ra^2}  + 
\frac{1}{\la r \pm \theta|y| \ra^2}\right)dr
\leq  \pi C. 
\end{multline*}
This shows that \refeq(final) is an $L^1$-bounded 
kernel and we conclude the proof of the lemma. 
\edpf 

\begin{footnotesize}

\end{footnotesize}

\end{document}